\documentclass[letterpaper,12pt]{article}
\usepackage{array,epsfig,rotating,setspace,latexsym,amsmath,epsf,amssymb,bm}
\usepackage{cite,graphicx,authblk,multirow,color,caption,subcaption,theorem}
\usepackage[title]{appendix}

\newtheorem{Theo}{Theorem}

\newtheorem{Lem}{Lemma}
\newtheorem{Cor}{Corollary}

\allowdisplaybreaks

\title{Secrecy in MIMO Networks with No Eavesdropper CSIT\thanks{This work was supported by NSF Grants CNS 13-14733, CCF 14-22111, CCF 14-22129 and CNS 15-26608, and presented in part at the Allerton Conference, Monticello, IL, September 2016.}}

\author{Pritam Mukherjee \qquad Sennur Ulukus\\
	\normalsize Department of Electrical and Computer Engineering\\
	\normalsize University of Maryland, College Park, MD 20742\\
	\normalsize {\it pritamm@umd.edu} \qquad {\it ulukus@umd.edu}}

\def\Y{\mathbf{Y}}

\def\X{\mathbf{X}}
\def\Z{\mathbf{Z}}
\def\N{\mathbf{N}}
\def\H{\mathbf{H}}
\def\G{\mathbf{G}}
\def\P{\mathbf{P}}
\def\Q{\mathbf{Q}}

\def\v{\mathbf{v}}
\def\u{\mathbf{u}}
\def\barP{\bar{\P}}
\def\barQ{\bar{\Q}}
\def\barH{\bar{\H}}
\def\barG{\bar{\G}}
\def\barY{\bar{\Y}}
\def\barZ{\bar{\Z}}
\def\barN{\bar{\N}}
\allowdisplaybreaks
\begin{document}
	\maketitle
	\begin{abstract}
		We consider two fundamental multi-user channel models: the multiple-input multiple-output (MIMO) wiretap channel with one helper (WTH) and the MIMO multiple access wiretap channel (MAC-WT). In each case, the eavesdropper has $K$ antennas while the remaining terminals have $N$ antennas each. We consider a fast fading channel where the channel state information (CSI) of the legitimate receiver is available at the transmitters but no channel state information at the transmitters (CSIT) is available for the eavesdropper's channel. We determine the optimal sum secure degrees of freedom (s.d.o.f.) for each channel model for the regime $K\leq N$, and show that in this regime, the MAC-WT channel reduces to the WTH in the absence of eavesdropper CSIT. For the regime $N\leq K\leq 2N$, we obtain the optimal \emph{linear} s.d.o.f., and show that the MAC-WT channel and the WTH have the same optimal s.d.o.f.~when restricted to linear encoding strategies. In the absence of any such restrictions, we provide an upper bound for the sum s.d.o.f.~of the MAC-WT chanel in the regime $N\leq K\leq 2N$. Our results show that unlike in the single-input single-output (SISO) case, there is loss of s.d.o.f.~for even the WTH due to lack of eavesdropper CSIT when $K\geq N$.  
	\end{abstract}
	\section{Introduction}
	We consider two multi-user models: the multiple-input multiple-output (MIMO) wiretap channel with one helper (WTH) where the transmitter, the helper and the legitimate receiver have $N$ antennas each, and the eavesdropper has $K$ antennas; see Fig.~\ref{fig:wt_model}, and the MIMO multiple access wiretap channel (MAC-WT), where both transmitters and the legitimate receiver have $N$ antennas each and the eavesdropper has $K$ antennas; see Fig.~\ref{fig:mac_model}. In both cases, the channel is fast fading  and the channel gains vary in an independent and identically distributed (i.i.d.) fashion across the links and time. We consider the case when the eavesdropper's channel state information (CSI) is not available at the transmitters (no eavesdropper CSIT). Our goal in this paper is to investigate the optimal sum secure degrees of freedom (s.d.o.f.) of the MIMO WTH and the MIMO MAC-WT channel as a function of $N$ and $K$. 
	
	To that end, we provide an achievable scheme based on vector space alignment \cite{cadambe_jafar_interference}, that attains $\frac{1}{2}(2N-K)$ s.d.o.f.~for the WTH for all values of $0\leq K \leq 2N$. Note that when $K\leq N$, this value coincides with the optimal s.d.o.f.~for the WTH in the case where full eavesdropper CSIT is available. Therefore, for the regime $K\leq N$, there is no loss of s.d.o.f.~for the WTH due to the lack of eavesdropper CSIT. Further, the proposed scheme which does not require eavesdropper CSIT, is optimal. The achievable scheme for the WTH also suffices as an achievable scheme for the MAC-WT channel, since we can treat one of the transmitters as a helper and use time-sharing among the two transmitters. 
	
	To prove the optimality of the proposed scheme for the MAC-WT channel, we next provide a matching converse for the regime $K\leq N$. Besides using MIMO versions of the \emph{secrecy penalty lemma} and the \emph{role of a helper lemma} \cite{jianwei_ulukus_one_hop}, the converse proof relies on exploiting channel symmetry at the eavesdropper. Since the transmitters do not have the eavesdropper's CSIT, the output at the $K$ antennas of the eavesdropper are \emph{entropy symmetric} \cite{kobayashi_delayed_csit}, i.e., any two subsets of the antenna outputs have the same differential entropy, if the subsets are of equal size. Finally, we use a MIMO version of the \emph{least alignment lemma} \cite{aligned_image_sets_jafar,mimome_blindwiretap_delayedcsit}, which states that the differential entropy at the output of the terminal which does not provide CSIT is the greatest among terminals having equal number of antennas. Intuitively, this holds since no signal alignment is possible at the output of the terminal which does not provide CSIT. 
	
	The converse in the regime $K\leq N$ shows that the sum s.d.o.f.~cannot exceed $\frac{1}{2}(2N-K)$ for the MAC-WT channel. Note that a converse for the MAC-WT channel is valid for the WTH as well. Further, together with the achievable scheme, it shows that the optimal s.d.o.f.~for both the WTH and the MAC-WT channel in this regime is $\frac{1}{2}(2N-K)$; therefore, as in the SISO case \cite{pmukherjee_ulukus_mac_isit2015,pritam_one_hop}, which is a subset of this regime with $N=K=1$, the MAC-WT channel reduces to the WTH when the eavesdropper's CSIT is not available. Recalling that with full eavesdropper CSIT, the optimal sum s.d.o.f.~of the MAC-WT channel in this regime is $\min(N,\frac{2}{3}(2N-K))$ \cite{pritam_ulukus_alsilomar15,pritam_mimo_mac_fullcsi}, this also illustrates the loss of s.d.o.f.~for the MAC-WT channel due to the lack of eavesdropper's CSIT.  
	
	Next, we consider the regime $N\leq K \leq 2N$. In this regime, we provide an upper bound which shows that the sum s.d.o.f.~of the MAC-WT channel cannot be larger than $\min\left(\frac{N}{2},\frac{2N(2N-K)}{4N-K}\right)$. Noting that $\frac{2N(2N-K)}{4N-K} < (2N-K)$, we conclude that there will be loss of s.d.o.f.~due to lack of eavesdropper CSIT, even for the WTH, in the regime $\frac{4N}{3}\leq K \leq 2N$, where $\min\left(\frac{N}{2},2N-K\right)$ s.d.o.f.~is achievable with full eavesdropper CSIT \cite{nafea_yener,nafea_yener_journal}. 
	
	In order to further investigate the optimality of $\frac{1}{2}(2N-K)$ as the sum s.d.o.f.~for the MAC-WT channel in the regime $N\leq K \leq 2N$, we then restrict ourselves to \emph{linear} encoding strategies \cite{x_channel_linear_dof,wiretap_helper_delayed}, where the channel input of each antenna in every time slot is restricted to be a linear combination of some information symbols intended for the legitimate receiver and some artificial noise symbols to provide secrecy at the eavesdropper. We show that under this restriction to linear encoding schemes, the \emph{linear} sum s.d.o.f.~can be no larger than $\frac{1}{2}(2N-K)$. The key idea of the proof is that since no alignment is possible at the eavesdropper, the artificial noise symbols should asymptotically occupy the maximum number of dimensions available at the eavesdropper; consequently, the dimension of the linear signal space at the eavesdropper should be $Kn+o(n)$ in $n$ channel uses.    
	
	\emph{Related Work:} The MAC-WT channel is introduced by \cite{tekin-yener-it2,tekin_yener_mac2008}, where the technique of cooperative jamming is introduced to improve the rates achievable with Gaussian signaling. Reference \cite{ersen_ulukus_mac_2008} provides outer bounds and identifies cases where these outer bounds are within 0.5 bits per channel use of the rates achievable by Gaussian signaling. While the exact secrecy capacity remains unknown, the achievable rates in \cite{tekin-yener-it2,tekin_yener_mac2008,ersen_ulukus_mac_2008} all yield zero s.d.o.f. Positive s.d.o.f.~can be obtained by either structured signaling \cite{structured_codes_he_yener_2014_journal} or non-i.i.d.~Gaussian signaling \cite{bassily_ergodic_align}. The exact optimal sum s.d.o.f.~of the wiretap channel with $M$ helpers and the $K$-user MAC-WT channel are established to be $\frac{M}{M+1}$ and $\frac{K(K-1)}{K(K-1)+1}$, respectively in \cite{jianwei_ulukus_one_hop}, when full eavesdropper's CSIT is available. References \cite{pmukherjee_ulukus_mac_isit2015,jianwei_ulukus_helper_2013,pritam_one_hop} show that without eavesdropper's CSIT, the optimal s.d.o.f.~for the wiretap channel with $M$ helpers is still $\frac{M}{M+1}$, while the optimal sum s.d.o.f.~of the $K$-user MAC-WT channel decreases to $\frac{K-1}{K}$. The two-user MIMO WTH, with full eavesdropper CSIT is considered in \cite{nafea_yener_initial,nafea_yener,nafea_yener_journal}, and the optimal s.d.o.f.~is determined for the case when the transmitter and the receiver each has $N$ antennas, the helper has $K$ antennas and the eavesdropper has $M$ antennas. References \cite{pritam_mimo_mac_fullcsi,pritam_ulukus_alsilomar15,pritam_ulukus_icc16} determine the optimal sum s.d.o.f.~for the two user MIMO MAC-WT channel when each transmitter and the receiver have $N$ antennas while the eavesdropper has $K$ antennas, and full eavesdropper CSIT is available. 
	
	A related line of research investigates the MIMO wiretap channel, the MIMO MAC-WT, and the MIMO broadcast channel with an \emph{arbitrarily varying} eavesdropper \cite{XHe_Yener2014,XHe_Yener2013,broadcast_he_yener}, when the eavesdropper CSIT is not available. The eavesdropper's channel is assumed to be arbitrary, without any assumptions on its distribution, and security is guaranteed for \emph{every} realization of the eavesdropper's channel. This models an exceptionally strong eavesdropper, which may control its own channel in an adversarial manner.  When $K\geq N$, the eavesdropper's channel realizations may be exactly equal to the legitimate user's channel realizations, and therefore, the optimal sum s.d.o.f.~is zero in this regime for both the MAC-WT and the WTH. When $K\leq N$, and the channel matrices to the legitimate receiver are full rank, the optimal sum s.d.o.f.~is $N-K$ for both the MAC-WT and the WTH. On the other hand, in our model, the entries in the eavesdropper's channel matrices are drawn from a known distribution, though the realizations are not known at the transmitters. We show that, with this mild assumption, strictly positive s.d.o.f.~can be achieved even when $K\geq N$. Further, the s.d.o.f.~achieved in our case when $K\leq N$ is strictly larger than the optimal s.d.o.f.~of $N-K$ for the case with an arbitrarily varying eavesdropper.

	\section{System Model}
	In this paper, we consider two fundamental channel models: the MIMO WTH and the MIMO MAC-WT. In each case, we assume that the channel gains are non-zero and are drawn from a common continuous distribution with bounded support in an i.i.d.~fashion in each channel use. The common continuous distribution is known at all the terminals in the system. We assume no eavesdropper CSIT, that is, the channel gains to the eavesdropper are not available at any transmitter. In the following three subsections we describe each channel model and provide the relevant definitions. 
	
	\subsection{Wiretap Channel with a Helper}
	\begin{figure}[t]
		\centering
		\vspace{10 pt}
		\includegraphics[height=200 pt]{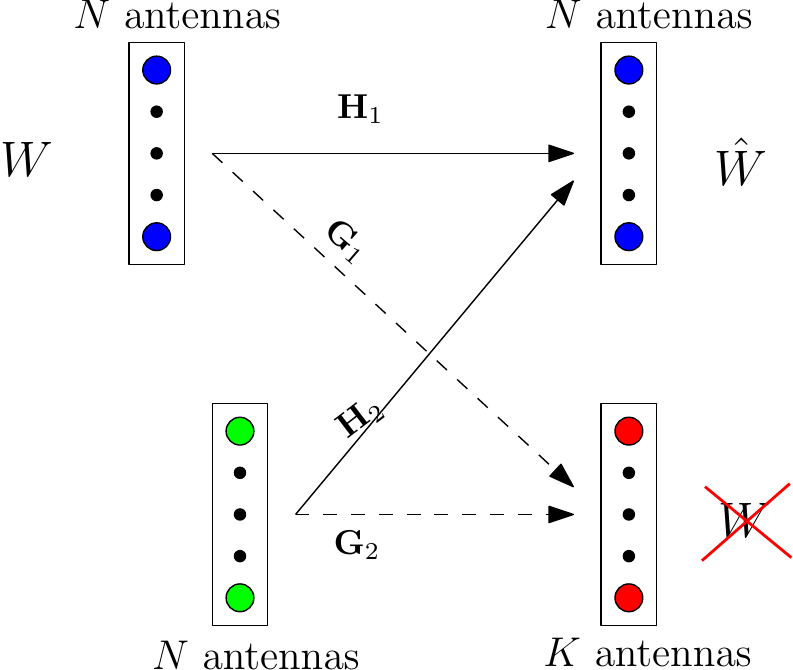}
		\caption{Wiretap channel with a helper (WTH).}
		\label{fig:wt_model}
	\end{figure}
	
	The MIMO WTH, see Fig.~\ref{fig:wt_model}, is described by,
	\begin{align}
	\Y(t) =& \H_1(t)\X_1(t) + \H_2(t)\X_2(t) + \N_1(t)\\
	\Z(t) =& \G_1(t)\X_1(t) + \G_2(t)\X_2(t) + \N_2(t)
	\end{align}
	where $\X_1(t)$ and $\X_2(t)$ are $N$ dimensional column vectors denoting the input of the legitimate transmitter and the helper, respectively, $\Y(t)$ is an $N$ dimensional vector denoting the legitimate receiver's channel output, and $\Z(t)$ is a $K$ dimensional vector denoting the eavesdropper's channel output, at time $t$. In addition, $\N_1(t)$ and $\N_2(t)$ are $N$ and $K$ dimensional white Gaussian noise vectors, respectively, with $\N_1 \sim \mathcal{N}(\mathbf{0},\mathbf{I}_{N})$ and  $\N_2 \sim \mathcal{N}(\mathbf{0},\mathbf{I}_{K})$, where $\mathbf{I}_N$ denotes the $N\times N$ identity matrix. Here, $\H_i(t)$ and $\G_i(t)$ are the $N\times N$ and $K\times N$ channel matrices from transmitter $i$ to the legitimate receiver and the eavesdropper, respectively, at time $t$. The entries of $\H_i(t)$ and $\G_i(t)$ are drawn from a fixed continuous distribution with bounded support in an i.i.d.~fashion at every time slot $t$. We assume that the channel matrices at the legitimate receiver, $\H_i(t)$, are known with full precision at all terminals, at time $t$. However, the channel matrices to the eavesdropper, $\G_i(t)$ are not known at any transmitter. All channel inputs satisfy the average power constraint $E[\lVert \X_i(t)\rVert^2]\leq P,\; i=1,2$, where $\lVert \X \rVert$ denotes the Euclidean (or spectral) norm of the vector (or matrix) $\X$.
	
	The transmitter wishes to send a message $W$, uniformly distributed in $\mathcal{W}$, securely to the legitimate receiver in the presence of the eavesdropper. A secure rate $R$, with $R = \frac{\log|\mathcal{W}|}{n}$ is achievable if there exists a sequence of codes which satisfy the reliability constraints at the legitimate receiver, namely, $\mbox{Pr}[ W\neq \hat{W}] \leq \epsilon_{n}$, for $i=1,2$, and the secrecy constraint, namely,
	\begin{align}
	\frac{1}{n} I(W;\Z^n) \leq \epsilon_n
	\end{align}
	where $\epsilon_n \rightarrow 0$ as $n \rightarrow \infty$. An s.d.o.f.~$d$ is said to be achievable if a rate $R$ is achievable with 
	\begin{align}
	d = \lim\limits_{P\rightarrow \infty} \frac{R}{\frac{1}{2}\log P}
	\end{align}
	
	\subsection{Multiple Access Wiretap Channel}
		\begin{figure}[t]
			\centering
			\vspace{10 pt}
			\includegraphics[height=200 pt]{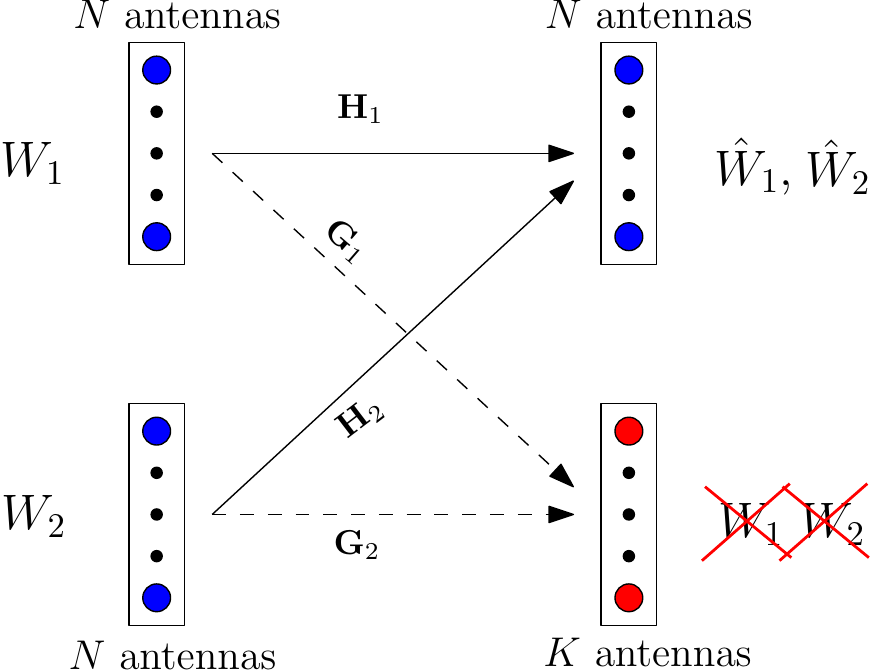}
			\caption{Multiple access wiretap channel (MAC-WT).}
			\label{fig:mac_model}
		\end{figure}
	The two-user MIMO MAC-WT, see Fig.~\ref{fig:mac_model}, is as follows:
	\begin{align}
	\Y(t) =& \H_1(t)\X_1(t) + \H_2(t)\X_2(t) + \N_1(t)\\
	\Z(t) =& \G_1(t)\X_1(t) + \G_2(t)\X_2(t) + \N_2(t)
	\end{align}
	where $\X_i(t)$ is an $N$ dimensional column vector denoting the $i$th user's channel input, $\Y(t)$ is an $N$ dimensional vector denoting the legitimate receiver's channel output, and $\Z(t)$ is a $K$ dimensional vector denoting the eavesdropper's channel output, at time $t$. In addition, $\N_1(t)$ and $\N_2(t)$ are $N$ and $K$ dimensional white Gaussian noise vectors, respectively, with $\N_1 \sim \mathcal{N}(\mathbf{0},\mathbf{I}_{N})$ and  $\N_2 \sim \mathcal{N}(\mathbf{0},\mathbf{I}_{K})$, where $\mathbf{I}_N$ denotes the $N\times N$ identity matrix. Here, $\H_i(t)$ and $\G_i(t)$ are the $N\times N$ and $K\times N$ channel matrices from transmitter $i$ to the legitimate receiver and the eavesdropper, respectively, at time $t$. The entries of $\H_i(t)$ and $\G_i(t)$ are drawn from a fixed continuous distribution with bounded support in an i.i.d.~fashion at every time slot $t$. We assume that the channel matrices to the legitimate receiver, $\H_i(t)$, are known with full precision at all terminals, at time $t$. However, the channel matrices to the eavesdropper, $\G_i(t)$, are not available at the transmitters. All channel inputs satisfy the average power constraint $E[\lVert \X_i(t)\rVert^2]\leq P,\; i=1,2$.
	
	Transmitter $i$ wishes to send a message $W_i$, uniformly distributed in $\mathcal{W}_i$, securely to the legitimate receiver in the presence of the eavesdropper. A secure rate pair $(R_{1},R_{2})$, with $R_{i} = \frac{\log|\mathcal{W}_i|}{n}$ is achievable if there exists a sequence of codes which satisfy the reliability constraints at the legitimate receiver, namely, $\mbox{Pr}[ W_{i}\neq \hat{W}_{i}] \leq \epsilon_{n}$, for $i=1,2$, and the secrecy constraint, namely,
	\begin{align}
	\frac{1}{n} I(W_{1},W_2;\Z^n) \leq \epsilon_n
	\end{align}
	where $\epsilon_n \rightarrow 0$ as $n \rightarrow \infty$. An s.d.o.f.~pair  $\left(d_1, d_2 \right) $ is said to be achievable if a rate pair $\left(R_1,R_2 \right) $ is achievable with 
	\begin{align}
	d_i = \lim\limits_{P\rightarrow \infty} \frac{R_i}{\frac{1}{2}\log P}
	\end{align}
	The sum s.d.o.f.~$d_s$ is the largest achievable $d_1+d_2$.
	
	\subsection{A Linear Secure Degrees of Freedom Perspective}\label{sec:lin_strat}
	
	In this paper, we also consider \emph{linear} coding strategies as defined in \cite{bresler_tse_mimo_interference,x_channel_linear_dof}. In such cases, the degrees of freedom simply represents the dimension of the linear subspace of transmitted signals. 
	
	When we focus on linear coding schemes, we consider a communication scheme of blocklength $n$, where transmitter $i$ wishes to send $m_i(n)$ \emph{information} symbols $\v_i\in \mathbb{R}^{m_i(n)}$ to the legitimate receiver reliably and securely. In case of the WTH, $m_2(n) = 0$. Each information symbol is a zero-mean Gaussian random variable with variance $\alpha P$, where $\alpha$ is a constant chosen to ensure that the power constraints are satisfied at each transmitter. In addition to the information symbols, transmitter $i$ can use $n_i(n)$ artificial noise symbols, $\u_i \in \mathbb{R}^{n_i(n)}$ each of which is a zero-mean Gaussian random variable with variance $\alpha P$. These artificial noise symbols need not be decoded at the receiver; instead they drown out the information symbols at the eavesdropper for security.  
	
	At each time $t$, the information symbols $\v_i$ at transmitter $i$ are modulated by a precoding matrix $\P_i(t) \in \mathbb{R}^{N\times m_i(n)}$, while the artificial noise symbols $\u_i$ are modulated using a precoding matrix $\Q_i(t)\in \mathbb{R}^{N\times n_i(n)}$. Since the channel gains $\H_i(t),\, i=1,2$ are known at both transmitters at time $t$, the precoding matrices $\P_i(t)$ and $\Q_i(t)$ can each depend on $\left\lbrace \H_1(k),\H_2(k), k=1,\ldots,t\right\rbrace $.  However, since the channel gains $\G_i(t)$ are not available at any transmitter, $\P_i$ and $\Q_i$ are independent of $\left\lbrace \G_i(t), t=1,\ldots,n\right\rbrace $.
	
	At time $t$, transmitter $i$ sends a linear combination of the information and the artificial noise symbols:
	\begin{align}
	\X_i(t) = \P_i(t)\v_i + \Q_i(t)\u_i\label{eq:channel_input}
	\end{align} 
	The channel outputs at time $t$ are, therefore,
	\begin{align}
	\Y(t) =& \H_1(t)\P_1(t)\v_1 + \H_2(t)\P_2(t)\v_2 + \H_1(t)\Q_1(t)\u_1 + \H_2(t)\Q_2(t)\u_2 + \N_1(t)\\
	\Z(t) =& \G_1(t)\P_1(t)\v_1 + \G_2(t)\P_2(t)\v_2+ \G_1(t)\Q_1(t)\u_1 + \G_2(t)\Q_2(t)\u_2 + \N_2(t)
	\end{align}
	Now letting $\barP_i = [\P_i(1), \ldots, \P_i(n)]^T$, $\barQ_i = [\Q_i(1),\ldots, \Q_i(n)]$, we can compactly write the channel outputs as
		\begin{align}
	  \barY = \barH_1\barP_1\v_1 + \barH_2\barP_2\v_2 + \barH_1\barQ_1\u_1 + \barH_2\barQ_2\u_2 + \barN_1 \label{eq:output1}\\
		\barZ = \barG_1\barP_1\v_1 + \barG_2\barP_2\v_2 + \barG_1\barQ_1\u_1 + \barG_2\barQ_2\u_2 + \barN_2\label{eq:output2}
		\end{align}
	where $\barH_i$ and $\barG_i$ are the $Nn\times Nn$ and $Kn \times Nn$ block diagonal matrices
	\begin{align}
	\barH_i =& \left[\begin{array}{cccc}
	\H_i(1) & \mathbf{0} & \ldots & \mathbf{0}\\
	\mathbf{0} &\H_i(2) & \ldots &\mathbf{0}\\
	\vdots & \vdots & \ddots & \vdots\\
	\mathbf{0} & \mathbf{0} & \ldots & \H_i(n)
	\end{array} \right],\qquad
		\barG_i =& \left[\begin{array}{cccc}
		\G_i(1) & \mathbf{0} & \ldots & \mathbf{0}\\
		\mathbf{0} &\G_i(2) & \ldots &\mathbf{0}\\
		\vdots & \vdots & \ddots & \vdots\\
		\mathbf{0} & \mathbf{0} & \ldots & \G_i(n)
		\end{array} \right]
	\end{align}
	and $\barN_i = [\N_i(1),\ldots, \N_i(n)]^T$ for $i=1,2$.
	
	At the legitimate receiver, the interference subspace is 
	\begin{align}
	 \mathcal{I}_B = \mbox{colspan}([\barH_1\barQ_1, \barH_2\barQ_2])
	\end{align}
	Let $\mathcal{I}_B^c$ denote the orthogonal subspace of $\mathcal{I}_B$. If we ignore the additive Gaussian noise, i.e., in the high transmit power regime, the decodability of $\v_1$ and $\v_2$ at the legitimate receiver corresponds to the constraint that the projection of the subspace colspan$([\barH_1\barP_1, \barH_2\barP_2])$ onto $\mathcal{I}_B^c$ must have dimension $m_1(n)+m_2(n)$, i.e.,
	\begin{align}
\mbox{dim}\left(\mbox{Proj}_{\mathcal{I}_B^c} \mbox{colspan}\left([\barH_1\barP_1, \barH_2\barP_2] \right) \right) 
	&= \mbox{dim}\left(\mbox{colspan}\left([\barP_1]  \right) \right)+\mbox{dim}\left(\mbox{colspan}\left([\barP_2]  \right) \right) \nonumber\\&= m_1(n) + m_2(n) \label{eq:reliability}
	\end{align} 
	This can be rewritten as requiring that
	\begin{align}
	\mbox{rank}\left([\barH_1\barP_1, \barH_2\barP_2, \barH_1\barQ_1, \barH_2\barQ_2] \right) -  \mbox{rank}\left([ \barH_1\barQ_1, \barH_2\barQ_2] \right) =  m_1(n) + m_2(n) \label{eq:mod_reliability}
	\end{align} 
	
	On the other hand, at the eavesdropper, we require that 
	\begin{align}
	\lim_{n\rightarrow \infty} \frac{1}{n} \mbox{dim}\left(\mbox{Proj}_{\mathcal{I}_E^c} \mbox{colspan}\left([\barG_1\barP_1, \barG_2\barP_2] \right) \right) = 0,\, a.s. \label{eq:security}
	\end{align} 
	where $\mathcal{I}_E = \mbox{colspan}([\barG_1\barQ_1, \barG_2\barQ_2])$.
	
	The security requirement in \eqref{eq:security} can be reformulated as follows: Let $L(n)$ be the number of \emph{leakage dimensions} defined as \begin{align}
	L(n) =& \mbox{rank}\left([\barG_1\barP_1, \barG_2\barP_2, \barG_1\barQ_1, \barG_2\barQ_2] \right)  -  \mbox{rank}\left([ \barG_1\barQ_1, \barG_2\barQ_2] \right) 
	\end{align}
	Then, we want 
	\begin{align}
	\lim_{n\rightarrow \infty} \frac{L(n)}{n} = 0, \, a.s. \label{eq:mod_sec_condition}
	\end{align}
	In other words, we want the artificial noise symbols to occupy the full received signal space at the eavesdropper asymptotically. 
	
	The quantity $L(n)$ may be thought of as the evaluation of $\lim_{P\rightarrow \infty}\frac{I(\v_1,\v_2;\bar{\Z})}{\frac{1}{2}\log P}$ for the input-output relation stated in \eqref{eq:output2}. To see this, we use \cite[Lemma 1]{pritam_one_hop}, which we state here for completeness.	
	\begin{Lem}\label{lem:dif_entropy}
		Let $\mathbf{A}$ be an $M\times N$ dimensional matrix and let $\mathbf{X} = \left(X_1,\ldots,X_N\right)^T$ be a jointly Gaussian random vector with zero-mean and variance $P\mathbf{I}$. Also, let $\mathbf{N} = \left(N_1,\ldots, N_M\right)^T$ be a jointly Gaussian random vector with zero-mean and variance $\sigma^2\mathbf{I}$, independent of $\mathbf{X}$. If $r=\mbox{rank}(\mathbf{A})$, then,
		\begin{align}
		h(\mathbf{A}\mathbf{X} + \mathbf{N}) = r\left(\frac{1}{2}\log P\right) + o(\log P)
		\end{align}
	\end{Lem}
	
	Using Lemma \ref{lem:dif_entropy}, we have
	\begin{align}
	I(\v_1,\v_2;\bar{\Z}) =& h(\bar{\Z}) - h(\bar{\Z}|\v_1,\v_2)\\
	=& \left(\mbox{rank}\left([\barG_1\barP_1, \barG_2\barP_2, \barG_1\barQ_1, \barG_2\barQ_2] \right)  -  \mbox{rank}\left([ \barG_1\barQ_1, \barG_2\barQ_2] \right)\right) \left(\frac{1}{2}\log P\right)\nonumber\\& + o(\log P) \\
	=& L(n)\left(\frac{1}{2}\log P\right) + o(\log P) 
	\end{align}
	which implies 
	\begin{align}
	\lim_{P\rightarrow\infty} \frac{I(\v_1,\v_2;\bar{\Z})}{\frac{1}{2}\log P} = L(n)
	\end{align} 	
	In a similar way, the decodability requirement in \eqref{eq:mod_reliability} can be thought of as ensuring that $\lim_{P\rightarrow \infty}\frac{I(\v_1,\v_2;\bar{\Y})}{\frac{1}{2}\log P} = m_1(n)+ m_2(n)$, for the input output relation stated in \eqref{eq:output1}. 
	
	For the WTH, a \emph{linear} s.d.o.f.~$d^{lin}$ with  $d^{lin} = m_1(n)/n$ is said to be achievable if there exists  a sequence of precoding matrices $\barP_1, \barQ_1, \barQ_2$ such that both the reliability constraints in \eqref{eq:reliability} and the security constraints in \eqref{eq:security} are satisfied. 
	
	For the MAC-WT channel, a \emph{linear} s.d.o.f.~pair $(d_1^{lin},d_2^{lin})$, with $d_i^{lin} = m_i(n)/n$ is said to be achievable if there exists  a sequence of precoding matrices $\barP_i, \barQ_i$ such that both the reliability constraints in \eqref{eq:reliability} and the security constraints in \eqref{eq:security} are satisfied. The \emph{linear} sum s.d.o.f.~$d_s^{lin}$ is the supremum of $d_1^{lin}+d_2^{lin}$, such that the pair $(d_1^{lin},d_2^{lin})$ is achievable.

	\section{Main Results}
	The main result of this paper is the determination of the optimal linear sum s.d.o.f.~for the MIMO WTH and the MIMO MAC-WT channel. We have the following theorem.
	\begin{Theo}\label{theo:sumdof}
		For both the $N\times N\times N \times K$ WTH and the MAC-WT channel with no eavesdropper CSIT, the optimal linear sum s.d.o.f.~$d_s^{lin}$ is 
		\begin{align}
		d_s^{lin} = \max\left(\frac{1}{2}(2N-K),0\right)
		\end{align}
		for almost all channel gains. Further, without any linearity constraints, the optimal sum s.d.o.f.~$d_s$ is
		\begin{align}
		d_s \begin{cases}
		= \frac{1}{2}(2N-K),\quad& 0\leq K\leq N\\
		\leq \min\left(\frac{N}{2},\frac{2N(2N-K)}{4N-K}\right),\quad& N\leq K\leq 2N\\
		= 0, \quad& K\geq 2N
		\end{cases}
		\end{align}
	\end{Theo}
	
	We also have the following corollary.
	\begin{Cor}
		For the $N\times N\times N \times K$ MAC-WT channel with no eavesdropper CSIT, the linear s.d.o.f.~region is given by the set of all nonnegative pairs $(d_1^{lin},d_2^{lin})$ that satisfy,
		\begin{align}
		d_1^{lin}+d_2^{lin} = \frac{1}{2}(2N-K)
		\end{align}
	\end{Cor}
	
	The proof of the corollary follows from the observation that every point in the given region can be achieved by time sharing between the points $\left(\frac{1}{2}(2N-K),0 \right) $ and $\left(0,\frac{1}{2}(2N-K) \right) $, which can themselves be attained by treating the MAC-WT channel as a WTH. Also, no point outside the region is achievable since the sum s.d.o.f.~is bounded by $\frac{1}{2}(2N-K)$ from Theorem \ref{theo:sumdof}.
	
	\begin{figure}
		\centering
		\vspace{20 pt}
		\includegraphics[height=200 pt]{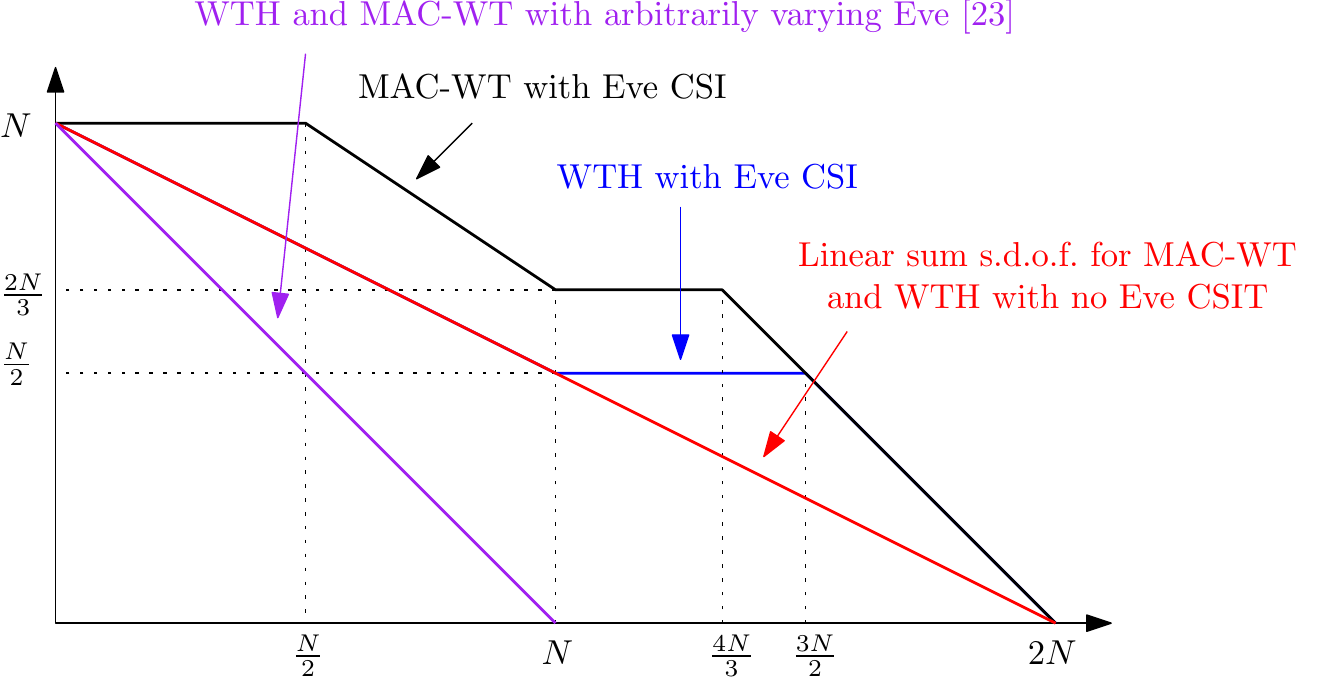}
		\caption{Sum s.d.o.f.~with number of eavesdropper antennas.}
		\label{fig:sdof_plot2}
	\end{figure}
	
\begin{figure}
			\centering
			\vspace{20 pt}
			\includegraphics[height=200 pt]{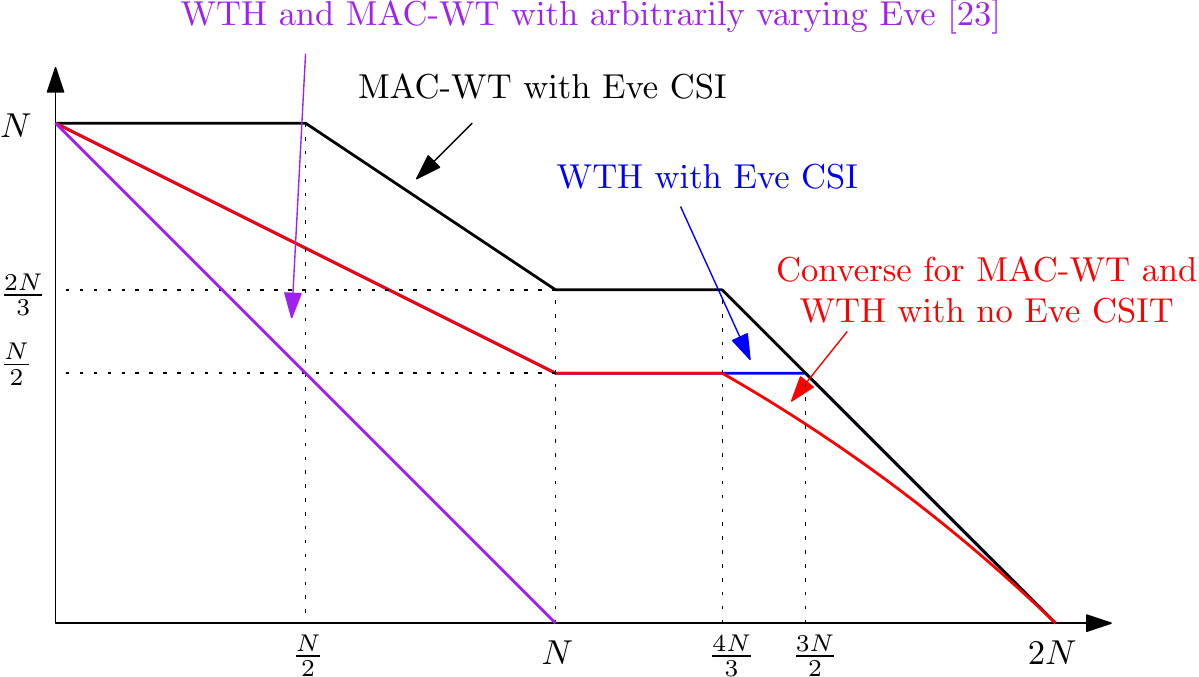}
			\caption{Converse for MAC-WT with no Eve CSIT.}
			\label{fig:sdof_plot1}
		\end{figure}

	Fig.~\ref{fig:sdof_plot2} shows the optimal linear sum s.d.o.f.~for the WTH and the MAC-WT channel with and without eavesdropper CSIT, while Fig.~\ref{fig:sdof_plot1} shows the general (without any linearity restrictions on the encoding schemes) upper bound on the sum s.d.o.f.~for the MAC-WT channel without eavesdropper CSIT. Similar to the SISO case \cite{pritam_one_hop}, as shown in Fig.~\ref{fig:sdof_plot2}, the MIMO MAC-WT channel reduces to the WTH when the eavesdropper CSIT is not available for the regime $0\leq K\leq N$, and at least from a linear s.d.o.f.~perspective in the regime $N\leq K\leq 2N$. However, unlike in the SISO case \cite{pritam_one_hop}, the linear s.d.o.f.~for the WTH decreases due to the lack of eavesdropper CSIT. Even without any linearity constraints, the optimal s.d.o.f.~for the WTH does decrease due to lack of eavesdropper CSIT, as can be seen from the general upper bound in Fig.~\ref{fig:sdof_plot1}, especially in the regime $\frac{4N}{3}\leq K \leq 2N$. Fig.~\ref{fig:sdof_plot2} and Fig.~\ref{fig:sdof_plot1} also show the optimal sum s.d.o.f.~for the WTH and the MAC-WT with an arbitrarily varying eavesdropper. When $K\geq N$, the optimal sum s.d.o.f.~is zero in this case since with $K=N$, the channel matrices of  eavesdropper channel may be made exactly equal to the channel matrices of the legitimate receiver's channel. When $K\leq N$ and the legitimate receiver's channel matrices are full rank, the optimal sum s.d.o.f.~is $N-K$ for both the MAC-WT and the WTH. Thus, we achieve a strictly larger sum s.d.o.f.~in our case where the entries of the eavesdropper's channel matrices are drawn from a continuous distribution and security is guaranteed on average, and not for every realization. 
	 
	\section{Proof of Theorem \ref{theo:sumdof}}
	In this section, we prove Theorem \ref{theo:sumdof} by providing an achievable scheme and a converse. Since Theorem \ref{theo:sumdof} implies that the WTH and the MAC-WT channel have the same linear sum s.d.o.f., we first note that it suffices to provide a linear achievable scheme for the WTH, since the MAC-WT channel can be treated as a WTH with time sharing between the users. Also, since any rate achievable for the WTH is achievable for the MAC-WT channel, a converse for the MAC-WT channel suffices as a converse for the WTH as well. Thus, in the following subsections, we provide an achievable scheme for the WTH and a converse for the MAC-WT channel.
	
	\subsection{Achievable Scheme for the WTH}
	In this section, we provide an achievable scheme for the WTH. In this scheme, the transmitter sends $(2N-K)$ information symbols reliably and securely to the legitimate receiver in two time slots, in order to achieve $\frac{1}{2}(2N-K)$ s.d.o.f. At each time slot, transmitter $1$ sends a linear combination of $(2N-K)$ information symbols $\v_1$ and $K$ artificial noise symbols $\u_1$ as in \eqref{eq:channel_input}.  Transmitter $2$ sends a linear combination of its $K$ artificial noise symbols $\u_2$. Since transmitter $2$ does not have any information symbols $\v_2$ for the WTH, there is no $\P_2$ in that case. The channel outputs can be written compactly as in \eqref{eq:output1}-\eqref{eq:output2} as:
	\begin{align}
		  \barY = \barH_1\barP_1\v_1  + \barH_1\barQ_1\u_1 + \barH_2\barQ_2\u_2 + \barN_1 \label{eq:scheme_output1}\\
		  \barZ = \barG_1\barP_1\v_1 + \barG_1\barQ_1\u_1 + \barG_2\barQ_2\u_2 + \barN_2\label{eq:scheme_output2}
	\end{align}
	where 
	\begin{align}
	\bar{\H}_i = \left[\begin{array}{cc}
	\H_i(1) & \mathbf{0}\\
	\mathbf{0} & \H_i(2)  	
		\end{array}\right], \qquad 
		\bar{\G}_i = \left[\begin{array}{cc}
		\G_i(1) & \mathbf{0}\\
		\mathbf{0} & \G_i(2)  	
		\end{array}\right]	
	\end{align}
	 It remains to choose the precoding matrices $\barP_1$, $\barQ_1$ and $\barQ_2$ appropriately. We make the following selection:
	 \begin{align}
	 \barQ_i = \barH_i^{-1}\barQ, \quad i=1,2
	 \end{align} 
	 where $\barQ$ is a $2N\times K$ matrix with rank $K$. Also choose $\barP_1$ to be a $2N\times (2N-K)$ matrix with rank $2N-K$, such that the matrix $[\barH_1\barP_1, \barQ]$ has rank $2N$. Note that this condition will be satisfied almost surely if the elements of $\barP_1$ and $\barQ$ are chosen from any continuous distribution in an i.i.d.~fashion. With this selection, the channel outputs are:
	 	\begin{align}
	 	\barY =& \barH_1\barP_1\v_1  + \barQ_1(\u_1 +\u_2) + \barN_1 \label{eq:final_output1}\\
	 	\barZ =& \barG_1\barP_1\v_1 + \barG_1\barH_1^{-1}\barQ\u_1 + \barG_2\barH_2^{-1}\barQ\u_2 + \barN_2\label{eq:final_output2}
	 	\end{align}
	 	
	 The decodability of $\v_1$ at the legitimate receiver in the high transmit power regime follows immediately since the matrix $[\barH_1\barP_1, \barQ]$ has rank $2N$ by our choice of $\barP_1$ and $\barQ$. On the other hand, the number of \emph{leakage dimensions} $L$ is
	 \begin{align}
	 L =& \mbox{rank}[\barG_1\barP_1,  \barG_1\barH_1^{-1}\barQ, \barG_2\barH_2^{-1}\barQ] - \mbox{rank}[\barG_1\barH_1^{-1}\barQ, \barG_2\barH_2^{-1}\barQ]\\
	 \leq& 2K - 2K \\
	 =& 0
	 \end{align}
	 where we have used the fact that for any full-rank $\barQ$ chosen independently of $\barG_1, \barG_2$, we have that $\mbox{rank}[\barG_1\barH_1^{-1}\barQ, \barG_2\barH_2^{-1}\barQ] = 2K$ for almost all channel realizations of $(\barG_1, \barG_2)$. This follows from the following lemma by noting that each row and each column of $\barG_i$ has at least one entry drawn from a continuous distribution in an i.i.d.~fashion and the matrices $\barH_i^{-1}\barQ$ for $i=1,2$ do not depend on the $\barG_i$s. 
	 
	 	\begin{Lem}\label{lem:main_lemma}
	 		Let $\P_1 \in \mathbb{R}^{N\times m_1}$ and $\P_2 \in \mathbb{R}^{N\times m_2}$ fixed matrices with ranks $p_1$ and $p_2$, respectively. Let $\G_1$ and $\G_2$ be $K\times N$ matrices whose each row and each column has at least one entry that is drawn from some continuous distribution in an i.i.d.~fashion, and the remaining elements are arbitrary but fixed. Then, almost surely,
	 		\begin{align}
	 		\text{\emph{rank}}[\G_1\P_1, \G_2\P_2] = \min\left(p_1+p_2,K\right) 
	 		\end{align}
	 	\end{Lem}
	The proof of this lemma is relegated to Appendix \ref{appendix:main_lemma_proof}.

	 Therefore, the security requirement in \eqref{eq:mod_sec_condition} is satisfied as well. This completes the achievable scheme. We remark here that using Lemma \ref{lem:dif_entropy}, it can be easily shown that $I(\v_1;\barY) = (2N-K)\left(\frac{1}{2}\log P\right)+ o(\log P)$ and $I(\v_1;\barZ) \leq o(\log P)$. An achievable rate for the wiretap channel over two channel uses, $R^{vec}$ is given by \cite{csiszar}
	 \begin{align}
	 R^{vec} =& I(\v_1;\barY) - I(\v_1;\barZ)\\
	  \geq& (2N-K)\left(\frac{1}{2}\log P\right)+ o(\log P)
	 \end{align} 
	 Therefore, the effective achievable secure rate is
	 \begin{align}
	 R \geq \frac{(2N-K)}{2}\left(\frac{1}{2}\log P\right)+ o(\log P)
	 \end{align}
	  which yields an s.d.o.f.~of $\frac{1}{2}(2N-K)$.
	 
	\subsection{Converse}
	In this section, we prove the converse for the MAC-WT channel. To that end, we consider two regimes of $K$. When $0\leq K \leq N$, we prove the converse for general transmission schemes without any restrictions of linearity. For the regime $N\leq K\leq 2N$, we prove the converse under the assumption of linear coding schemes only. We also provide a general upper bound in this regime which does not match the achievablity; nevertheless, it shows that there is loss in s.d.o.f.~for the WTH and the MAC-WT channel due to no eavesdropper CSIT. 
	
	\subsubsection{$0\leq K \leq N:$ Converse with No Restrictions}
	We wish to show that:
	\begin{align}
	d_1+d_2 \leq \frac{1}{2}(2N-K)
	\end{align}
    Let us first state three lemmas which are useful for the proof. 
    \begin{Lem}[Channel symmetry {\cite[Lemma 3]{kobayashi_delayed_csit}}]\label{lem:channel_symmetry}
    	Let $Z^K = \left\lbrace Z_1,\ldots, Z_K \right\rbrace $ be entropy symmetric, i.e., for any subsets $A$ and $B$ of $\left\lbrace 1,\ldots, K \right\rbrace $, with $|A| = |B|\leq K$, 
    	\begin{align}
    	h(\left\lbrace Z_i, i\in A\right\rbrace ) = h(\left\lbrace Z_i, i\in B\right\rbrace ) 
    	\end{align}
    	Then, for any $M\geq N$, the following holds:
    	\begin{align}
    	\frac{1}{N}h(Z^N) \geq \frac{1}{M}h(Z^M) 
    	\end{align}
    \end{Lem} 
    
    \begin{Lem}[Least alignment lemma {\cite[Lemma 3]{mimome_blindwiretap_delayedcsit}}]\label{lem:least_alignment}
    	Consider two receivers, each with $L$ antennas. Suppose the channel gains to receiver $2$ are not available at the transmitters. If $\Y$ and $\Z$ denote the channel outputs at receivers $1$ and $2$, respectively, we have
    	\begin{align}
    	h(\Z^n) \geq h(\Y^n) + n o(\log P)
    	\end{align}
    	
    \end{Lem}
    
    Combining the two lemmas, we have the following lemma.
    \begin{Lem}\label{lem:entropy_ineq}
    	For the $N\times N \times N\times K$ MIMO MAC-WT channel with no eavesdropper CSIT, with $K\leq N$
    	\begin{align}
    	h(\Z^n) \geq \frac{K}{N}h(\Y^n) + n o(\log P)
    	\end{align}
    \end{Lem}
    We relegate the proof of this lemma to Appendix \ref{appendix:proof_entropy_ineq}.
    
    Let us now proceed with the converse proof. As in \cite{jianwei_ulukus_one_hop,nafea_yener,nafea_yener_journal,pritam_ulukus_alsilomar15}, we define noisy versions of $\X_i$ as $\tilde{\X}_i = \X_i + \tilde{\N}_i$ where $\tilde{\N}_i \sim \mathcal{N}(\mathbf{0},\rho_i^2\mathbf{I}_N)$ with $\rho_i^2 < \min\left(\frac{1}{\lVert \H_i \rVert^2},\frac{1}{\lVert \G_i \rVert^2} \right)$. The \emph{secrecy penalty lemma} \cite{jianwei_ulukus_one_hop} can then be derived as
    \begin{align}
    n(R_1 + R_2) \leq& I(W_1,W_2;\Y^n| \Z^n) + n\epsilon\\
    \leq& h(\Y^n|\Z^n) + no(\log P) \\
    =& h(\Y^n,\Z^n) - h(\Z^n) +no(\log P)\\
    \leq& h(\tilde{\X}_1^n,\tilde{\X}_2^n) - h(\Z^n) + no(\log P)\\
    =& h(\tilde{\X}_1^n) + h(\tilde{\X}_2^n) - h(\Z^n) + no(\log P) \label{eq:sec_penalty_lemma1}   
    \end{align}  
    
    The \emph{role of a helper lemma} \cite{jianwei_ulukus_one_hop} also generalizes to the MIMO case as
    \begin{align}
    nR_1 \leq& I(\X_1^n;\Y^n)\\
    =& h(\Y^n) - h(\H_2^n\X_2^n + \N_1^n)\\
    \leq& h(\Y^n) - h(\tilde{\X}_2^n) + no(\log P) \label{eq:role_of_helper1}
    \end{align}
    By symmetry, we also have
    \begin{align}
        nR_2 \leq& h(\Y^n) - h(\tilde{\X}_1^n) + no(\log P) \label{eq:role_of_helper2}
    \end{align}    
    Adding \eqref{eq:sec_penalty_lemma1}, \eqref{eq:role_of_helper1} and \eqref{eq:role_of_helper2}, we have
    \begin{align}
    2n(R_1+R_2) \leq& 2h(\Y^n) - h(\Z^n) + n o(\log P)\\
    \leq& 2h(\Y^n) - \frac{K}{N}h(\Y^n)+ n o(\log P)\label{eq:using_lemma3}\\
    =&\frac{2N-K}{N}h(\Y^n) + n o(\log P)\\
    \leq& (2N-K)\left(\frac{n}{2}\log P \right) + n o(\log P) 
    \end{align}
    where \eqref{eq:using_lemma3} follows from Lemma \ref{lem:entropy_ineq} and we have used the fact that $h(\Y^n)\leq \frac{N}{2}\log P + n o(\log P)$. Therefore, we have, 
    \begin{align}
    R_1+R_2 \leq \frac{1}{2}(2N-K)\left(\frac{1}{2}\log P \right) + o(\log P)
    \end{align}
     Dividing by $\frac{1}{2}\log P$ and taking the limit $P\rightarrow \infty$, we have
     \begin{align}
     d_1+d_2 \leq \frac{1}{2}(2N-K)
     \end{align}
     which completes the proof of the converse for the regime $0\leq K\leq N$.
     
     \subsubsection{$N\leq K \leq 2N:$ Converse with Linear Coding Strategies}
     We begin with the following lemma.
     \begin{Lem}\label{lem:full_space_lemma}
     	For the $N\times N\times N\times K$ MAC-WT channel, and for any \emph{linear} achievable scheme satisfying both the reliability and security constraints, and also $d_1+d_2 > 0$, 
     	\begin{align}
  \lim_{n\rightarrow\infty} \frac{1}{n}\text{\emph{rank}}\left([\barG_1\barP_1, \barG_2\barP_2, \barG_1\barQ_1, \barG_2\barQ_2] \right)=      	\lim_{n\rightarrow\infty} \frac{1}{n}\text{\emph{rank}}\left([\barG_1\barQ_1, \barG_2\barQ_2] \right) = K
     	\end{align} 
     \end{Lem}
     We relegate the proof of this lemma to Appendix \ref{appendix:proof_full_space_lemma}.
     
     To proceed with the upper bound, first note that since strictly positive sum s.d.o.f.~is achievable for the MAC-WT channel using linear schemes, we can safely discard the case $d_1+d_2=0$ for the purpose of the converse. Therefore, from Lemma \ref{lem:full_space_lemma}, the rank of the vector space spanned by the output at the eavesdropper is $Kn+o(n)$, i.e.,
     \begin{align}
\lim_{n\rightarrow\infty} \frac{1}{n}\mbox{rank}\left([\barG_1\barP_1, \barG_2\barP_2, \barG_1\barQ_1, \barG_2\barQ_2] \right)=      	\lim_{n\rightarrow\infty} \frac{1}{n}\mbox{rank}\left([\barG_1\barQ_1, \barG_2\barQ_2] \right) = K
     \end{align}
     
     We have,
     \begin{align}
     m_1(n)+ m_2(n) =& \mbox{rank}\left([\barH_1\barP_1, \barH_2\barP_2, \barH_1\barQ_1, \barH_2\barQ_2] \right) - \mbox{rank}\left([ \barH_1\barQ_1, \barH_2\barQ_2] \right)\label{eq:decodability}\\
     \leq& \mbox{rank}\left([\barH_1\barP_1, \barH_2\barP_2, \barH_1\barQ_1, \barH_2\barQ_2] \right) - \mbox{rank}\left([ \barH_1\barQ_1, \barH_2\barQ_2] \right)\nonumber\\
     &- \mbox{rank}\left([\barG_1\barP_1, \barG_2\barP_2, \barG_1\barQ_1, \barG_2\barQ_2] \right) + \mbox{rank}\left([ \barG_1\barQ_1, \barG_2\barQ_2] \right) +o(n) \label{eq:security_cond}\\
     \leq& \mbox{rank}\left([\barH_1\barP_1, \barH_2\barP_2, \barH_1\barQ_1, \barH_2\barQ_2] \right) - \frac{1}{2}\mbox{rank}\left([ \barG_1\barQ_1, \barG_2\barQ_2] \right)\nonumber\\
     &- \mbox{rank}\left([\barG_1\barP_1, \barG_2\barP_2, \barG_1\barQ_1, \barG_2\barQ_2] \right) + \mbox{rank}\left([ \barG_1\barQ_1, \barG_2\barQ_2] \right) +o(n)\label{eq:key_step1}\\
     =& \mbox{rank}\left([\barH_1\barP_1, \barH_2\barP_2, \barH_1\barQ_1, \barH_2\barQ_2] \right) + \frac{1}{2}\mbox{rank}\left([ \barG_1\barQ_1, \barG_2\barQ_2] \right)\nonumber\\
          &- \mbox{rank}\left([\barG_1\barP_1, \barG_2\barP_2, \barG_1\barQ_1, \barG_2\barQ_2] \right) +o(n)\\
     \leq& \mbox{rank}\left([\barH_1\barP_1, \barH_2\barP_2, \barH_1\barQ_1, \barH_2\barQ_2] \right) + \frac{1}{2}\mbox{rank}\left([\barG_1\barP_1, \barG_2\barP_2, \barG_1\barQ_1, \barG_2\barQ_2] \right)\nonumber\\
     &- \mbox{rank}\left([\barG_1\barP_1, \barG_2\barP_2, \barG_1\barQ_1, \barG_2\barQ_2] \right) +o(n)\\      
     \leq& \mbox{rank}\left([\barH_1\barP_1, \barH_2\barP_2, \barH_1\barQ_1, \barH_2\barQ_2] \right)+o(n)\nonumber\\ &-\frac{1}{2}\mbox{rank}\left([\barG_1\barP_1, \barG_2\barP_2, \barG_1\barQ_1, \barG_2\barQ_2] \right) \\ 
     \leq& Nn - \frac{1}{2}Kn +o(n) \label{eq:from_lemma_5}\\
     =& \frac{(2N-K)n}{2} +o(n)
     \end{align}
     where \eqref{eq:decodability} follows from the decodability constraint, \eqref{eq:security_cond} follows from the secrecy constraint \eqref{eq:mod_sec_condition}, and \eqref{eq:key_step1} follows from the following:
     \begin{align}
2 \times \mbox{rank}\left([ \barH_1\barQ_1, \barH_2\barQ_2] \right) &\geq \mbox{rank}\left([ \barH_1\barQ_1] \right) +\mbox{rank}\left([ \barH_2\barQ_2] \right) \\
     &= \mbox{rank}\left([\barQ_1] \right) +\mbox{rank}\left([\barQ_2] \right) \\
     &= \mbox{rank}\left([ \barG_1\barQ_1] \right) +\mbox{rank}\left([ \barG_2\barQ_2] \right) \\
     &\geq \mbox{rank}\left([ \barG_1\barQ_1, \barG_2\barQ_2] \right)
     \end{align}
     The above equalities all hold almost surely since $\barH_i$ and $\barG_i$ are both full column rank almost surely. Finally, \eqref{eq:from_lemma_5} follows from Lemma \ref{lem:full_space_lemma}.
     
      Now dividing by $n$ and taking limit $n\rightarrow \infty$, we have
      \begin{align}
      d_1+d_2 \leq \frac{1}{2}(2N-K)
      \end{align}

      \subsubsection{$N\leq K \leq 2N:$ Converse with No Restrictions}  We have the following lemma.
      \begin{Lem}\label{lem:entropy_ineq2}
      	For the $N\times N\times N\times K$ MIMO MAC-WT channel with no eavesdropper CSIT, with $K\leq 2N$
      	\begin{align}
      	h(\Z^n) \geq \frac{K}{2N}h(\Y^n, \Z^n) + n o(\log P)
      	\end{align}
      \end{Lem}
      The proof of this lemma is relegated to Appendix \ref{appendix:proof_entropy_ineq2}. 
      
      Now we proceed as in the case of $0\leq K \leq N$ with the \emph{secrecy penalty lemma} \cite{jianwei_ulukus_one_hop}:
      \begin{align}
      n(R_1 + R_2) \leq& I(W_1,W_2;\Y^n| \Z^n) + n\epsilon\\
      \leq& h(\Y^n|\Z^n) + no(\log P)\\
      \leq& h(\Y^n,\Z^n) - h(\Z^n) +no(\log P)\\
      \leq& \left( 1-\frac{K}{2N}\right) h(\Y^n,\Z^n)+ n o(\log P)\\
      \leq& \left( 1-\frac{K}{2N}\right) h(\tilde{\X}_1^n,\tilde{\X}_2^n)+ n o(\log P)\\
      =& \frac{2N-K}{2N}\left(h(\tilde{\X}_1^n) + h(\tilde{\X}_2^n) \right)+ n o(\log P) \label{eq:sec_penalty_lemma2} 
      \end{align}
      
      The \emph{role of the helper lemma} \cite{jianwei_ulukus_one_hop} yields, for $i\neq j$:
          \begin{align}
          nR_i \leq& h(\Y^n) - h(\tilde{\X}_j^n) + n o(\log P) \label{eq:role_of_helper3}
          \end{align}
          
       Eliminating $h(\tilde{\X}_1^n)$ and $h(\tilde{\X}_2^n)$ using \eqref{eq:sec_penalty_lemma2} and \eqref{eq:role_of_helper3},
       \begin{align}
       n(R_1+R_2) \leq& \frac{2(2N-K)}{4N-K}h(\Y^n) + n o(\log P)\\
       \leq& \frac{2N(2N-K)}{4N-K}\left(\frac{n}{2}\log P \right) + n o(\log P) 
       \end{align}
       Dividing by $n$ and letting $n\rightarrow \infty$, we have
       \begin{align}
       R_1+R_2 \leq \frac{2N(2N-K)}{4N-K}\left(\frac{1}{2}\log P \right) +  o(\log P) 
       \end{align}
       Now dividing by $\frac{1}{2}\log P$ and letting $P \rightarrow \infty$,
       \begin{align}
       d_1+d_2 \leq \frac{2N(2N-K)}{4N-K}
       \end{align}
       Also, $d_1+d_2\leq \frac{N}{2}$, since $\frac{N}{2}$ is the optimal sum s.d.o.f.~when $K=N$, and the sum s.d.o.f.~is non-increasing in $K$.

	\section{Conclusions}
	In this paper, we considered two fundamental multi-user channel models: the MIMO WTH and the MIMO MAC-WT channel. In each case, the eavesdropper has $K$ antennas while the remaining terminals have $N$ antennas. We assumed that the CSIT of the legitimate receiver is available but no eavesdropper CSIT is available. We determined the optimal sum s.d.o.f.~for each channel model for the regime $K\leq N$, and showed that in this regime, the MAC-WT channel reduces to the WTH in the absence of eavesdropper CSIT. For the regime $N\leq K\leq 2N$, we obtained the optimal \emph{linear} s.d.o.f., and showed that the MAC-WT channel and the WTH have the same optimal s.d.o.f.~when restricted to linear encoding strategies. In the absence of any such restrictions, we provided an upper bound for the sum s.d.o.f.~of the MAC-WT channel in the regime $N\leq K\leq 2N$. Our results showed that unlike in the SISO case, there is loss of s.d.o.f.~for even the WTH due to lack of eavesdropper CSIT, especially when $K\geq N$. 
	
	\begin{appendices}
	\section{Proof of Lemma \ref{lem:main_lemma}}\label{appendix:main_lemma_proof}
			First note when $N\leq K$, $\G_i$s have full column rank almost surely. Therefore,
			\begin{align}
			\mbox{rank}[\G_i\P_i] = \mbox{rank}[\P_i] = p_i
			\end{align}
			almost surely. On the other hand, when $N\geq K$, we have 
			\begin{align}
			\mbox{rank}[\G_i\P_i] \geq \mbox{rank}[\G_i\hat{\P}_i]	
			\end{align}
			where $\hat{\P}_i$ is a $N\times p_i$ submatrix of $\P_i$ with full column rank. Let $\bar{p}_i = \min(K,p_i)$. Now, the determinant of any $\bar{p}_i\times \bar{p}_i$ submatrix of $\G_i\hat{\P}_i$ is a multi-variate polynomial of the random entries of $\G_i$ and is zero for only finitely many realizations. Therefore, $\G_i\hat{\P}_i$ has rank $\bar{p}_i$. Note that when $N\leq K$, $\bar{p}_i = p_i$ is satisfied trivially.

			 Therefore, there exists a set $I_i \subseteq \left\lbrace 1,\ldots,m_i \right\rbrace $ such that $|I_i| = \bar{p}_i$ and 
			the collection of column vectors $\mathbf{C}_i=\left\lbrace \mathbf{c}_{ij}, j\in I_i\right\rbrace $ are linearly independent, where $\mathbf{c}_{ij}$ denotes the $j$th column of $\G_i\P_i$. Clearly,
			\begin{align}
			\mbox{rank}[\G_1\P_1, \G_2\P_2] \geq \mbox{rank}[\mathbf{C}_1, \mathbf{C}_2] \label{eq:lem_step_initial}
			\end{align}
			The matrix $[\mathbf{C}_1, \mathbf{C}_2]$ is a $K\times (\bar{p}_1+\bar{p}_2)$ matrix. Now, if $K\leq \bar{p}_1+\bar{p}_2$, consider any $K\times K$ submatrix of $[\mathbf{C}_1, \mathbf{C}_2]$. The determinant of this submatrix is a multi-variate polynomial function of the random entries of $\G_1$ and $\G_2$, and therefore, the determinant can be zero for only finitely many realizations, corresponding to the roots of the multi-variate polynomial function. Note that this is true if each row and each column of $\barG_i$ has at least one random entry. Also, the polynomial function is not identically zero. Therefore, 
			\begin{align}
			\mbox{rank}[\mathbf{C}_1, \mathbf{C}_2] = K \label{eq:lem_step1}
			\end{align}
			
			On the other hand, if $K\geq \bar{p}_1+\bar{p}_2$, we can consider a $(\bar{p}_1+\bar{p}_2)\times (\bar{p}_1+\bar{p}_2)$ submatrix of $[\mathbf{C}_1, \mathbf{C}_2]$, and using a similar argument, we can claim that 
			\begin{align}
			\mbox{rank}[\mathbf{C}_1, \mathbf{C}_2] = \bar{p}_1+\bar{p}_2 \label{eq:lem_step2}
			\end{align}
			
			Combining \eqref{eq:lem_step_initial}, \eqref{eq:lem_step1} and \eqref{eq:lem_step2}, we have 
			\begin{align}
			\mbox{rank}[\G_1\P_1, \G_2\P_2] 
			\geq& \min\left(\bar{p}_1+\bar{p}_2,K\right) \\
			=& \min\left(\min(p_1,K)+ \min(p_2,K), K\right)\\
			=& \min \left(\min(p_1+p_2,K+p_1, K+p_2, 2K), K\right)\\
			=& \min\left(p_1+p_2,K\right)\label{eq:lbound}
			\end{align}
			
			On the other hand,
			\begin{align}
			\mbox{rank}[\G_1\P_1, \G_2\P_2] \leq& \mbox{rank}[\G_1\P_1] +  \mbox{rank}[\G_2\P_2]\\
			\leq& \min(\mbox{rank}[\G_1],p_1) + \min(\mbox{rank}[\G_2],p_2)\label{eq:matrix_mult_ineq}\\
			=& \min(N,K,p_1) + \min(N,K,p_2)\label{eq:full_rank_G}\\
			=& \min(K,p_1) + \min(K,p_2)\label{eq:rank_ineq} 
			\end{align}
			where \eqref{eq:matrix_mult_ineq} follows since $\mbox{rank}[AB] \leq \min(\mbox{rank}[A], \mbox{rank}[B])$, \eqref{eq:full_rank_G} follows since $\G_i$ is full rank almost surely, and \eqref{eq:rank_ineq} follows since $N\geq p_i$. 			
			Finally, it trivially holds that $K\geq \mbox{rank}[\G_1\P_1, \G_2\P_2]$.  Therefore, we have,
			\begin{align}
			\mbox{rank}[\G_1\P_1, \G_2\P_2] \leq& \min(K,\min(K,p_1) + \min(K,p_2) )\\
			=& \min(K,p_1+p_2)\label{eq:ubound}
			\end{align}
			Combining \eqref{eq:lbound} and \eqref{eq:ubound} completes the proof of the lemma.
			
		\section{Proof of Lemma \ref{lem:entropy_ineq}}\label{appendix:proof_entropy_ineq}
		Note that $K\leq N$. Consider $N-K$ additional outputs $\hat{\Z}$ at the eavesdropper as:
		\begin{align}
		\hat{\Z}(t) = \hat{\G}_1(t)\X_1(t) + \hat{\G}_2(t) \X_2(t) + \hat{\N}_2(t)
		\end{align}
		where each $\hat{\G}_i$ is a $(N-K)\times N$ matrix whose entries are drawn in an i.i.d.~fashion from the same continuous distribution as the entries of $\G_i$, and the entries of $\hat{\N}_2$ are i.i.d.~zero-mean unit-variance Gaussian noise. Assume that the $\hat{\G}_i$s are unavailable at the transmitters.. The enhanced output $\bar{\Z}(t) = (\Z(t),\hat{\Z}(t))$ is clearly entropy symmetric. Using Lemma \ref{lem:channel_symmetry}, we have
		\begin{align}
			h(\Z^n) \geq \frac{K}{N} h(\bar{\Z}^n) \label{eq:entropy_ineq_step1}
		\end{align}
		Now, since the $\G_i$s and $\hat{\G}_i$s are not available at the transmitters, using Lemma \ref{lem:least_alignment}, we have
		\begin{align}
		h(\bar{\Z}^n) \geq h(\Y^n) + n o(\log P)\label{eq:entropy_ineq_step2}
		\end{align}
		Combining \eqref{eq:entropy_ineq_step1} and \eqref{eq:entropy_ineq_step2}, we get the desired result that
		\begin{align}
		h(\Z^n) \geq \frac{K}{N}h(\Y^n) + n o(\log P)
		\end{align}

	 \section{Proof of Lemma \ref{lem:full_space_lemma}}\label{appendix:proof_full_space_lemma}
	 Since $d_1+d_2>0$, without loss of generality, assume $d_1>0$. We wish to prove that 	 
	 \begin{align}
	 &\hspace{-20pt}\lim_{n\rightarrow\infty} \frac{1}{n}\mbox{rank}\left([\barG_1\barP_1, \barG_2\barP_2, \barG_1\barQ_1, \barG_2\barQ_2] \right)=      	\lim_{n\rightarrow\infty} \frac{1}{n}\mbox{rank}\left([\barG_1\barQ_1, \barG_2\barQ_2] \right) = K
	 \end{align}
	 For the sake of contradiction, suppose $\lim_{n\rightarrow\infty} \frac{1}{n}\mbox{rank}\left([\barG_1\barQ_1, \barG_2\barQ_2] \right) < K$. We have
	 \begin{align}
	 \mbox{rank}\left([\barG_1\barP_1, \barG_2\barP_2, \barG_1\barQ_1, \barG_2\barQ_2] \right)&\geq \mbox{rank}\left([\barG_1\barP_1, \barG_1\barQ_1, \barG_2\barQ_2] \right)  \\
	 &= \mbox{rank}\left([\barG_1[\barP_1, \barQ_1], \barG_2\barQ_2] \right)\\
	 &\geq \min\left(\mbox{rank}\left([\barP_1, \barQ_1] \right) + \mbox{rank}\left([\barQ_2] \right) , Kn \right)\label{eq:using_main_lem}\\
	 &= \min\left(\mbox{rank}\left([\barP_1]\right) + \mbox{rank}\left( [\barQ_1] \right) + \mbox{rank}\left([\barQ_2] \right) , Kn \right)\label{eq:decoding}\\ 
	 &= \min\left(m_1(n)+ \mbox{rank}\left( [\barG_1\barQ_1] \right) + \mbox{rank}\left([\barG_2\barQ_2] \right) , Kn \right) \label{eq:full_rankG}\\
	 &\geq \min\left(m_1(n)+ \mbox{rank}\left( [\barG_1\barQ_1, \barG_2\barQ_2] \right), Kn \right)	  
	 \end{align}
	 where \eqref{eq:using_main_lem} follows from Lemma \ref{lem:main_lemma}, \eqref{eq:decoding} follows from the decodability requirement, and \eqref{eq:full_rankG} follows almost surely since $\barG_i$ is full column rank almost surely as long as $K>N$. 	 
	 Therefore, 
	 \begin{align}
	 \lim_{n\rightarrow \infty}\frac{1}{n}\mbox{rank}\left([\barG_1\barP_1, \barG_2\barP_2, \barG_1\barQ_1, \barG_2\barQ_2] \right) &\geq \min\left( d_1+ \lim_{n\rightarrow\infty}\frac{1}{n}\mbox{rank}\left( [\barG_1\barQ_1, \barG_2\barQ_2] \right), K \right) \\
	 &>  \lim_{n\rightarrow\infty}\frac{1}{n}\mbox{rank}\left( [\barG_1\barQ_1, \barG_2\barQ_2] \right)
	 \end{align}
	 which contradicts the security requirement in \eqref{eq:mod_sec_condition}.

	\section{Proof of Lemma \ref{lem:entropy_ineq2}}\label{appendix:proof_entropy_ineq2}
	 Consider $2N-K$ additional outputs $\hat{\Z}$ at the eavesdropper:
	 \begin{align}
	 \hat{\Z}(t) = \hat{\G}_1(t)\X_1(t) + \hat{\G}_2(t) \X_2(t) + \hat{\N}_2(t)
	 \end{align}
	 where each $\hat{\G}_i$ is a $(2N-K)\times N$ matrix whose entries are drawn in an i.i.d.~fashion from the same continuous distribution as the entries of $\G_i$, and the entries of $\hat{\N}_2$ are i.i.d.~zero-mean unit-variance Gaussian noise. Assume that the $\hat{\G}_i$s are not available at the transmitters either. Then, the enhanced output $\bar{\Z}(t) = (\Z(t),\hat{\Z}(t))$ is clearly entropy symmetric. Therefore, using Lemma \ref{lem:channel_symmetry}, we have
	 \begin{align}
	 h(\Z^n) \geq \frac{K}{2N} h(\bar{\Z}^n) \label{eq:entropy_ineq_step11}
	 \end{align}
	 Now, given $\bar{\Z}^n$, we can decode both inputs $\X_1^n$ and $\X_2^n$ to within noise variance, and therefore, also $\Y^n$ and $\Z^n$. Thus, we have
	 \begin{align}
	 h(\bar{\Z}^n) \geq h(\Y^n,\Z^n) + n o(\log P)\label{eq:entropy_ineq_step21}
	 \end{align}
	 Combining \eqref{eq:entropy_ineq_step11} and \eqref{eq:entropy_ineq_step21}, we get the desired result that
	 \begin{align}
	 h(\Z^n) \geq \frac{K}{2N}h(\Y^n,\Z^n) + n o(\log P)
	 \end{align}

	\end{appendices}
	\bibliographystyle{unsrt}
	\bibliography{references}
\end{document}